\documentstyle[12pt]{article}

\catcode`\@=11
\long\def\@makefntext#1{
\protect\noindent \hbox to 3.2pt {\hskip-.9pt
$^{{\ninerm\@thefnmark}}$\hfil}#1\hfill}		

\def\@makefnmark{\hbox to 0pt{$^{\@thefnmark}$\hss}}  

\def\ps@myheadings{\let\@mkboth\@gobbletwo
\def\@oddhead{\hbox{}
\rightmark\hfil\ninerm\thepage}
\def\@oddfoot{}\def\@evenhead{\ninerm\thepage\hfil
\leftmark\hbox{}}\def\@evenfoot{}
\def\sectionmark##1{}\def\subsectionmark##1{}}

\renewcommand{\thefootnote}{\fnsymbol{footnote}}

\newcounter{sectionc}\newcounter{subsectionc}\newcounter{subsubsectionc}
\renewcommand{\section}[1] {\vspace*{0.6cm}\addtocounter{sectionc}{1}
\setcounter{subsectionc}{0}\setcounter{subsubsectionc}{0}\noindent
	{\normalsize\bf\thesectionc. #1}\par\vspace*{0.4cm}}
\renewcommand{\subsection}[1] {\vspace*{0.6cm}\addtocounter{subsectionc}{1}
	\setcounter{subsubsectionc}{0}\noindent
	{\normalsize\it\thesectionc.\thesubsectionc. #1}\par\vspace*{0.4cm}}
\renewcommand{\subsubsection}[1]
{\vspace*{0.6cm}\addtocounter{subsubsectionc}{1}
	\noindent {\normalsize\rm\thesectionc.\thesubsectionc.\thesubsubsectionc.
	#1}\par\vspace*{0.4cm}}

\newcounter{appendixc}
\newcounter{subappendixc}[appendixc]
\newcounter{subsubappendixc}[subappendixc]

\renewcommand{\appendix}[1] {\vspace*{0.6cm}
        \refstepcounter{appendixc}
        \setcounter{figure}{0}
        \setcounter{table}{0}
        \setcounter{equation}{0}
        \renewcommand{\thefigure}{\Alph{appendixc}.\arabic{figure}}
        \renewcommand{\thetable}{\Alph{appendixc}.\arabic{table}}
        \renewcommand{\theappendixc}{\Alph{appendixc}}
        \renewcommand{\theequation}{\Alph{appendixc}.\arabic{equation}}
        \noindent{\bf Appendix \theappendixc #1}\par\vspace*{0.4cm}}

\def\abstracts#1{{

\centering{\begin{minipage}{12.2truecm}\footnotesize\baselineskip=12pt\noindent
	\centerline{\footnotesize ABSTRACT}\vspace*{0.3cm}
	\parindent=0pt #1
	\end{minipage}}\par}}

\renewenvironment{thebibliography}[1]
	{\begin{list}{\arabic{enumi}.}
	{\usecounter{enumi}\setlength{\parsep}{0pt}
\setlength{\leftmargin 1.25cm}{\rightmargin 0pt}
	 \setlength{\itemsep}{0pt} \settowidth
	{\labelwidth}{#1.}\sloppy}}{\end{list}}

\newcounter{itemlistc}
\newcounter{romanlistc}
\newcounter{alphlistc}
\newcounter{arabiclistc}

\newcommand{\fcaption}[1]{
        \refstepcounter{figure}
        \setbox\@tempboxa = \hbox{\footnotesize Fig.~\thefigure. #1}
        \ifdim \wd\@tempboxa > 6in
           {\begin{center}
        \parbox{6in}{\footnotesize\baselineskip=12pt Fig.~\thefigure. #1}
            \end{center}}
        \else
             {\begin{center}
             {\footnotesize Fig.~\thefigure. #1}
              \end{center}}
        \fi}

\newcommand{\tcaption}[1]{
        \refstepcounter{table}
        \setbox\@tempboxa = \hbox{\footnotesize Table~\thetable. #1}
        \ifdim \wd\@tempboxa > 6in
           {\begin{center}
        \parbox{6in}{\footnotesize\baselineskip=12pt Table~\thetable. #1}
            \end{center}}
        \else
             {\begin{center}
             {\footnotesize Table~\thetable. #1}
              \end{center}}
        \fi}

\def\@citex[#1]#2{\if@filesw\immediate\write\@auxout
	{\string\citation{#2}}\fi
\def\@citea{}\@cite{\@for\@citeb:=#2\do
	{\@citea\def\@citea{,}\@ifundefined
	{b@\@citeb}{{\bf ?}\@warning
	{Citation `\@citeb' on page \thepage \space undefined}}
	{\csname b@\@citeb\endcsname}}}{#1}}

\newif\if@cghi
\def\cite{\@cghitrue\@ifnextchar [{\@tempswatrue
	\@citex}{\@tempswafalse\@citex[]}}
\def\citelow{\@cghifalse\@ifnextchar [{\@tempswatrue
	\@citex}{\@tempswafalse\@citex[]}}
\def\@cite#1#2{{$\null^{#1}$\if@tempswa\typeout
	{IJCGA warning: optional citation argument
	ignored: `#2'} \fi}}

 1
 1
 1

\font\ninerm=cmr9

\newcommand{\be}{\begin{equation}}
\newcommand{\bea}{\begin{eqnarray}}
\newcommand{\ba}{\begin{array}}
\newcommand{\bean}{\begin{eqnarray*}}
\newcommand{\ee}{\end{equation}}
\newcommand{\eea}{\end{eqnarray}}
\newcommand{\ea}{\end{array}}
\newcommand{\eean}{\end{eqnarray*}}

\newcommand{\eg}{\mbox{$\gamma$}}

\def\Sq{{\sigma_{\mu\nu}q^{\nu}}}
\def\Sm{{\frac{i\Sq}{2m_{t}}}}

\begin{document}

\centerline{\normalsize\bf RECENT DEVELOPMENTS}
\baselineskip=22pt
\centerline{\normalsize\bf IN THE PINCH TECHNIQUE
\footnote{To appear in the proceedings of the Ringberg Workshop on
"Perspectives for electroweak interactions in $e^+e^-$  collisions"
hosted by the Max Planck Institut, at the Ringberg Castle, M\"unich,
February 5-8, 1995. Edited by B. Kniehl.} }
\baselineskip=16pt

\vfill
\vspace*{0.6cm}
\centerline{\footnotesize JOANNIS PAPAVASSILIOU}
\baselineskip=13pt
\centerline{\footnotesize\it Department of Physics, New York University,
4 Washington Place}
\baselineskip=12pt
\centerline{\footnotesize\it NY, NY 10003, USA}
\centerline{\footnotesize E-mail: papavass@mafalda.physics.nyu.edu}

\vspace*{0.9cm}
\abstracts{Some of the most important theoretical
and phenomenological aspects of the pinch technique
are presented, and several recent developments
are briefly reviewed.}

\normalsize\baselineskip=15pt
\setcounter{footnote}{0}
\renewcommand{\thefootnote}{\alph{footnote}}

\section{General considerations}
The pinch technique (PT)
 is an algorithm that allows the construction
of modified gauge independent
(g.i.) off-shell $n$-point functions, through the
order by order rearrangement of
Feynman graphs contributing to a certain physical
and therefore ostensibly g.i. process,
such as an S-matrix element or a Wilson loop.\cite{CornMarseille}
The PT was originally introduced
in an attempt
to device a
consistent truncation scheme for the
Schwinger-Dyson equations (SDE)
that govern the dynamics of gauge theories.
These equations are inherently non-perturbative and
 could in principle provide important
information about
a plethora of phenomena in non-Abelian gauge theories not
captured by perturbation theory. In practice however, one is
severely limited in exploiting them, mainly
because  they
 constitute an
 infinite set of coupled non-linear integral equations. Even though the
need for a truncation scheme is evident, particular care is needed for
respecting the crucial property of gauge invariance. Indeed, the SDE
are conventionally built out of gauge dependent Green's functions.
Since the mechanism of gauge cancellation is very subtle and involves a
delicate consiracy of terms
coming from all orders,
 a casual truncation of the series
 often gives rise to gauge dependent approximations for ostensibly
g.i. quantities.
  The PT attempts to address this problem in its root, namely
the building blocks of the SDE. According to this approach, the
Feynman graphs contributing to a given gauge invariant process are
rearranged
into new propagators and vertices where the gauge dependence has been
reduced to an absolute minimum, namely
that of the free gluon propagator.
The proper self-energy of the new propagator and the new vertices are
themselves g.i. and as it turns out so are the SDE
governing these new Green's functions. These new SDE are in
general more complicated than the usual ones because of the presence of
extra terms which enforce gauge invariance. Nonetheless, it is possible
to truncate them, usually by keeping only a few terms of a dressed loop
expansion, and maintain exact gauge invariance, while at the same time
accommodating non-perturbative effects.
One very important aspect of gauge invariance in the context of SDE
is that the Green's functions defined via the
PT satisfy {\it tree-level} Ward identities.This feature
is very important since it enables the cancellation of the final gauge
dependences stemming from the free parts of the gluon propagators
entering in the expressions for the SDE.

The systematic derivation of such a series for
QCD has been the focal
 point of extensive research.
 In a ghost free gauge,
the usual SDE for quarkless QCD
are build out of three basic quantities;
 the gluon propagator $\Delta$ , the three gluon
vertex  $\Gamma_{3}$, and the four gluon vertex $\Gamma_{4}$.
One then considers the effective potential $\Omega$,\cite{CJT}
a functional of the three fundamental
Green's functions,\cite{Mike and Dick}
 and then extremizes independently the
 variations of
 $\Omega(\Delta,\Gamma_{3},\Gamma{4})$ with respect to
$\Delta$, $\Gamma_{3}$, and $\Gamma_{4}$, e.g.
$\frac{\delta\Omega}{\delta\Delta} = 0$,
$\frac{\delta\Omega}{\delta\Gamma_3} = 0$, and
$\frac{\delta\Omega}{\delta\Gamma_4} = 0$.
 The resulting
 expressions will be the corresponding SDE for
$\Delta$, $\Gamma_{3}$ and $\Gamma_{4}$.
 In such a picture the
solutions to the SDE will in general be gauge dependent
in a non-trivial way. If one could solve the entire
renormalized set of SDE
and then substitute the resulting gauge dependent solutions
$\bar{\Delta}$, ${\bar{\Gamma}}_{3}$ and ${\bar{\Gamma}}_{4}$
 back into
$\Omega(\Delta,\Gamma_{3},\Gamma_{4})$
 and calculate its value,
$\Omega(\bar{\Delta},{\bar{\Gamma}}_{3},{\bar{\Gamma}}_{4})$,
the final answer would be g.i.,
since $\Omega$ is a physical quantity (vacuum energy).
 The way
this gauge cancellations would manifest
themselves is complicated
and involves non-trivial mixing of all orders.
However, since solving
he entire series is practically impossible, some form of truncation
is necessary.
The minimum requirement for
any such truncation scheme must be that the
solutions of the truncated SDE, when substituted into
$\Omega$, should still preserve its gauge invariance. Unfortunately
this is not the case if one truncates the series without a particular
guiding principle.
The alternative approach that has been proposed
{}~\cite{Schrodinger}
is to demand from the beginning that
the effective potential
 $\Omega(\hat{\Delta},{\hat{\Gamma}}_{3},{\hat{\Gamma}}_{4})$, as
well as the individual expressions for the self-energy
$\hat{d}$, for ${\hat{\Gamma}}_{3}$ and for ${\hat{\Gamma}}_{4}$, should
be g.i. order by order in the dressed loop expansion
(we use hats to indicate that these expressions are in general different
from their conventionally derived unhatted counterparts).
Assuming that
$\hat{d}$, ${\hat{\Gamma}}_{3}$ and ${\hat{\Gamma}}_{4}$ are individually
g.i. is not sufficient however to guarantee the order by
order gauge independence of $\Omega$, because there is a residual
dependence on the gauge fixing parameter coming from the free part of the
propagators $\hat{\Delta}$ entering in the expression for $\Omega$.
 The necessary and sufficient condition for the order by order
cancellation of the residual gauge dependence is that the renormalized
self energy
 ${\hat{\Pi}}_{\mu\nu}$
is transverse, e.g.
\begin{equation}
q^{\mu}{\hat{\Pi}}_{\mu\nu} = 0~,
\label{FirstOne}
\end{equation}
order by order in the dressed expansion.
 It turns out that Eq(\ref{FirstOne}) can be satisfied as long as
$\hat{d}$, ${\hat{\Gamma}}_{3}$ and
${\hat{\Gamma}}_{4}$ satisfy the following Ward identities:

\begin{equation}
q_{1}^{\mu}\hat{\Gamma}_{\mu\nu\alpha}(q_{1},q_{2},q_{3})=
  t_{\nu\alpha}(q_{2})\hat{d}^{-1}(q_{2}) -
 t_{\nu\alpha}(q_{3})\hat{d}^{-1}(q_{3})~,
\label{WI1}
\end{equation}
and
\begin{equation}
q_{1}^{\mu}\hat{\Gamma}_{\mu\nu\alpha\beta}^{abcd} =
f_{abp}\hat{\Gamma}_{\nu\alpha\beta}^{cdp}(q_1+q_2,q_3,q_4) + c.p.~,
\label{WI2}
\end{equation}
with
$t_{\mu\nu}=q^{2}g_{\mu\nu} - q_{\mu}q_{\nu}$,
$\hat{d}^{-1}(q)=q^{2}-\hat{\Pi}(q)$,
$f^{abc}$ the structure constants of the gauge group,
and the abbreviation
c.p. in the r.h.s. stands for cyclic permutations.
If Eq(\ref{WI1}) and Eq(\ref{WI2}) are
satisfied, than $\Omega$ is manifestly
g.i. order by order in the dressed loop expansion
 and so are the SDE generated
after its variation.
In particular,
one should extremize independently the
variations of $\Omega(\hat{d},\hat{\Gamma}_{3},\hat{\Gamma}_{4})$
with respect to $\hat{d}$, $\hat{\Gamma}_{3}$, and $\hat{\Gamma}_{4}$,
e.g.
$\frac{\delta\Omega}{\delta\hat{\Delta}} = 0$,
$\frac{\delta\Omega}{\delta {\hat{\Gamma}}_{3}} = 0$ and
$\frac{\delta\Omega}{\delta {\hat{\Gamma}}_{4}} = 0$,
imposing Eq(\ref{WI2}) as an additional constraint.
Once solved these equations will give rise to g.i.
$\hat{d}$, $\hat{\Gamma}_{3}$, and $\hat{\Gamma}_{4}$.
Although this program has been layed out conceptually, its practical
implementation is as yet incomplete. One thing is certain however:
if Green's functions with the properties described above can arise out
of a self-consistent treatment of QCD, one should be able to construct
Green's functions with the same properties at the level of ordinary
perturbation theory after appropriate rearrangement of Feynman graphs.
The PT accomplishes this task by providing the systematic algorithm needed
to recover the desired Green's functions order by order in perturbation
theory. So, g.i.
two, three, and four- gluon vertices have already been
constructed via the PT at one-loop, and
they satisfy the Ward identities
of Eq(\ref{FirstOne})-Eq(\ref{WI2}).

\section{The pinch technique}
The simplest example that demonstrates how the PT works is the gluon
two point function.\cite{Bib}
Consider the $S$-matrix
element $T$ for the elastic scattering
such as $q_{1}{\bar{q}}_{2}\rightarrow q_{1}{\bar{q}}_{2}$,
where $q_{1}$,$q_{2}$ are two
on-shell test quarks with masses
$m_{1}$ and $m_{2}$.
To any order in perturbation
theory $T$ is independent
of the gauge fixing parameter $\xi$.
On the other hand, as an explicit calculation shows,
the conventionally defined proper self-energy
depends on $\xi$. At the one loop level this dependence is canceled by
contributions from other graphs, which,
at first glance, do not seem to be
propagator-like.
That this cancellation must occur and can be employed to define a
g.i. self-energy, is evident from the decomposition:
\begin{equation}
T(s,t,m_{1},m_{2})= T_{0}(t,\xi) +
T_{1}(t,m_{1},\xi)+T_{2}(t,m_{2},\xi)
+T_{3}(s,t,m_{1},m_{2},\xi)~,
\label{S-matrix}
\end{equation}
where the function $T_{0}(t,\xi)$ depends
kinematically only on the Mandelstam variable
$t=-({\hat{p}}_{1}-p_{1})^{2}=-q^2$,
 and not on $s=(p_{1}+p_{2})^{2}$ or on the
external masses.
Typically, self-energy, vertex, and box diagrams
contribute to $T_{0}$, $T_{1}$, $T_{2}$, and $T_{3}$, respectively.
Such contributions are $\xi$ dependent, in general. However, as the sum
$T(s,t,m_{1},m_{2})$ is g.i., it is easy to show that
Eq(\ref{S-matrix}) can be recast in the form
\begin{equation}
T(s,t,m_{1},m_{2})=
{\hat{T}}_{0}(t) + {\hat{T}}_{1}(t,m_{1})+
{\hat{T}}_{2}(t,m_{2})+
{\hat{T}}_{3}(s,t,m_{1},m_{2})~,
\label{S2-matrix}
\end{equation}
where the ${\hat{T}}_{i}$ ($i=0,1,2,3$) are {\sl individually}
$\xi$-independent.
  The propagator-like parts of vertex and box graphs
which enforce the gauge independence of $T_{0}(t)$,
 are called pinch parts.
They emerge every time a gluon propagator or an elementary
three-gluon vertex contributes a longitudinal $k_{\mu}$ to the original
graph's numerator. The action of such a term is
to trigger an elementary
Ward identity of the form
$\not\hspace*{-1.0mm}{k} =
(\not\hspace*{-1.0mm}{p}+\not\hspace*{-1.0mm}{k}-m)-
(\not\hspace*{-1.0mm}{p}-m)$
when it gets contracted with a $\gamma$ matrix.
The first term removes (pinches out) the
internal fermion propagator,
whereas the second vanishes on shell.
{}From the g.i. functions ${\hat{T}}_{i}$ ($i=1,2,3$)
one may now extract a g.i. effective gluon ($G$) self-energy
${\hat{\Pi}}_{\mu\nu}(q)$, g.i. $Gq_{i}{\bar{q}}_{i}$ vertices
${\hat{\Gamma}}_{\mu}^{(i)}$, and a g.i. box
$\hat{B}$, in the following way:
\begin{eqnarray}
&{\hat{T}}_{0}=g^{2}{\bar{u}}_{1}\gamma^{\mu}u_{1}
[(\frac{1}{q^{2}}){\hat{\Pi}}_{\mu\nu}(q)(\frac{1}{q^{2}})]
{\bar{u}}_{2}\gamma^{\nu}u_{2}~,\nonumber \\
&{\hat{T}}_{1}=g^{2}
{\bar{u}}_{1}{\hat{\Gamma}}_{\nu}^{(1)}u_{1}
(\frac{1}{q^{2}}){\bar{u}}_{2}\gamma^{\nu}u_{2}~,~~~~~~~~~~~~~~\\
&{\hat{T}}_{2}=g^{2}
{\bar{u}}_{1}\gamma^{\mu}u_{1}
(\frac{1}{q^{2}}){\bar{u}}_{2}
{\hat{\Gamma}}_{\nu}^{(2)}u_{2}~,~~~~~~~~~~~~~~\nonumber\\
&{\hat{T}}_{3}=\hat{B}~,~~~~~~~~~~~~~~~~~~~~~~~~~
{}~~~~~~~~~~~~~\nonumber
\label{DefOne}
\end{eqnarray}
where $u_{i}$ are the external spinors, and $g$ is the gauge coupling.
Since all hatted quantities in the above formula are g.i., their explicit
form may be calculated using any value of
the gauge-fixing parameter $\xi$, as long as one
properly
identifies and allots all relevant pinch contributions. The
choice $\xi=1$ simplifies the calculations significantly, since
it eliminates the longitudinal part of the gluon propagator.
Therefore, for $\xi=1$ the pinch contributions originate only
from momenta carried by
the elementary three-gluon vertex
The one-loop expression for ${\hat{\Pi}}_{\mu\nu}(q)$
is given by \cite{Bib} :
\begin{equation}
{\hat{\Pi}}_{\mu\nu}(q)= {\Pi}_{\mu\nu}^{(\xi=1)}(q)+
t_{\mu\nu}{{\Pi}^{P}}(q)~,
\label{Prop}
\end{equation}
and
\begin{equation}
{{\Pi}^{P}}(q)=-2ic_{a}g^2
\int_{n}\frac{1}{k^2(k+q)^2}~,
\label{som}
\end{equation}
where $\int_{n}\equiv\int \frac{d^{n}k}{{(2\pi)}^{n}}$ is the
dimensionally regularized loop integral, and $c_{a}$ is the
quadratic Casimir operator for the adjoint representation
[for $SU(N)$, $c_{a}=N$]
After integration and renormalization we find
\begin{equation}
{{\Pi}^{P}}(q) = -2c_{a}(\frac{g^2}{16\pi^{2}})
\ln(\frac{-q^{2}}{\mu^{2}})]~.
\end{equation}
Adding this to the Feynman-gauge proper self-energy
\begin{equation}
{\Pi}_{\mu\nu}^{(\xi=1)}(q)=-[\frac{5}{3}
c_{a}(\frac{g^2}{16\pi^{2}})\ln(\frac{-q^{2}}{\mu^{2}})]
t_{\mu\nu}~,
\end{equation}
we find for ${\hat{\Pi}}_{\mu\nu}(q)$
\begin{equation}
{\hat{\Pi}}_{\mu\nu}(q)=-bg^2\ln(\frac{-q^{2}}{\mu^{2}})t_{\mu\nu}~,
\label{RunnCoupl}
\end{equation}
where $b=\frac{11c_{a}}{48\pi^{2}}$ is the coefficient of
$-g^{3}$ in the usual $\beta$ function.

This procedure can be extended to an arbitrary $n$-point function;
of particular physical interest are the g.i. three and
four point functions
${\hat{\Gamma}}_{\mu\nu\alpha}$~\cite{C&P} and
$\hat{\Gamma}_{\mu\nu\alpha\beta}$.\cite{4g}
Finally, the generalization of the PT to the case of
non-conserved external currents is technically more involved,
but conceptually straightforward.\cite{Bernd}

\section{The current algebra formulation of the pinch technique}

We now present an alternative
formulation of the PT introduced
in the context of the standard model.\cite{D&S} In this
approach the interaction of gauge bosons with external fermions
is expressed in terms of
current correlation functions,\cite{Another Sirlin}
i.e. matrix elements of Fourier transforms
of time-ordered products of current operators.
This is particularly economical because these amplitudes automatically
include several closely related Feynman diagrams. When one of the current
operators is contracted with the appropriate four-momentum, a Ward identity
is triggered. The pinch part is then identified with the contributions
involving the equal-time commutators in the Ward identities, and therefore
involve amplitudes in which the number of current operators has been
decreased by one or more. A basic ingredient in this formulation are the
following equal-time commutators;
\begin{eqnarray}
&\delta(x_0-y_0)[J^{0}_{W}(x),J^{\mu}_{Z}(y)]=
 c^{2}J^{\mu}_{W}(x)\delta^{4}(x-y)~,\nonumber\\
&\delta(x_0-y_0)[J^{0}_{W}(x),J^{\mu\dagger}_{W}(y)]=
 - J^{\mu}_{3}(x)\delta^{4}(x-y)~,\\
&\delta(x_0-y_0)[J^{0}_{W}(x),J^{\mu}_{\gamma}(y)]=
 J^{\mu}_{W}(x)\delta^{4}(x-y)~,~~\nonumber\\
&\delta(x_0-y_0)[J^{0}_{V}(x),J^{\mu}_{V^{'}}(y)]= 0~,~~~~~~~
{}~~~~~~~~~~~~~\nonumber
\label{Commut2}
\end{eqnarray}
where $J_{3}^{\mu}\equiv 2(J_{Z}^{\mu}+s^{2}J_{\gamma}^{\mu})$
and $V,V^{'} \in \{ \gamma,Z \}$.
To demonstrate the method with an example, consider
the vertex $\Gamma_{\mu}$, where now the gauge
particles in the loop are $W$ instead of
gluons and the incoming and outgoing fermions are massless.
It can be written as follows (with $\xi=1$):
\begin{equation}
\Gamma_{\mu}=\int \frac{d^{4}k}{{2\pi}^4}
\Gamma_{\mu\alpha\beta}(q,k,-k-q)\int d^{4}x e^{ikx}
<f|T^{*}[J^{\alpha\dagger}_{W}(x)J^{\beta}_{W}(0)]|i>~.
\label{Papous}
\end{equation}
When an appropriate momentum, say $k_{\alpha}$,
 from the vertex is pushed into the integral over
$dx$, it gets transformed into a covariant derivative
 $\frac{d}{dx_{\alpha}}$ acting on the time ordered product
$<f|T^{*}[J^{\alpha\dagger}_{W}(x)J^{\beta}_{W}(0)]|i>$.
After using current
conservation and differentiating the
$\theta$-function terms, implicit in the definition of
the $T^{*}$ product, we end up with the left-hand side
 of the second of Eq(\ref{Commut2}).
So, the contribution of each such term is proportional to the
matrix
element of a single current operator,
 namely $<f|J_{3}^{\mu}|i>$; this
is precisely the pinch part. Calling $\Gamma_{\mu}^{P}$
 the total pinch contribution from the
$\Gamma_{\mu}$ of Eq(\ref{Papous}), we find that
\begin{equation}
\Gamma_{\mu}^{P}= -g^{3}cI_{WW}(Q^2)<f|J_{3}^{\mu}|i>~,
\label{PinchPapou}
\end{equation}
where
\begin{equation}
I_{ij}(q)= i\int (\frac{d^{4}k}{2\pi^{4}})\frac{1}
{(k^{2}-M_{i}^{2})[{(k+q)}^{2}-M_{j}^{2}]}~.
\label{IntegralIWW}
\end{equation}
Obviously, the integral in Eq(\ref{IntegralIWW}) is the generalization
of the QCD expression Eq(\ref{som})
to the case of massive gauge bosons.

\section{Phenomenological applications}
In this section we present some of the most important
phenomenological applications of the PT.

\subsection{Neutrino electromagnetic form factor}
It has been known since the early days of gauge theories with
spontaneous symmetry breaking that both the electric and
magnetic form factors of fermions, $F_{1}(q^{2})$ and
$F_{2}(q^{2})$, respectively, defined by
\begin{equation}
\Gamma_{\mu}=\gamma_{\mu}F_{1}(q^{2})+
\frac{i}{2m}\sigma_{\mu\nu}q^{\nu}F_{2}(q^{2}),
\label{magmom}
\end{equation}
with
$\sigma_{\mu\nu}\equiv \frac{i}{2}[\gamma_{\mu},\gamma_{\nu}]$
are {\it gauge dependent} for general values of the momentum
transfer $q^{2}$. It is only at $q^{2}=0$ when the gauge dependence
drops so that $F_{1}(0)$ can be identified with the fermion charge,
and $F_{2}(0)$ with the anomalous magnetic moment.
In the context of the standard model the effective electromagnetic
form factor $F(q^{2})$ of the neutrino has been a long-standing
puzzle. It has been argued \cite{Bardeen} that the neutrino
mean-square radius $<r^{2}>$ and $F(q^{2})$ are related by
\begin{equation}
<r^{2}>=6 \frac{dF(q^{2})}{dq^{2}}|_{q^{2}=0}~~,
\label{radius}
\end{equation}
but it was soon realized that the conventional definition of
$F(q^{2})$ would give rise to gauge-dependent and divergent
expressions for $<r^{2}>$. This, of course, comes as no surprise.
There is indeed no a priori reason why even if $F(q^{2})$ is
g.i. at $q^{2}=0$, its derivative will be also.

The root of the problem lies in the fact that, although everyone agrees
that the Feynman diagrams are just convenient visualizations of a
complex underlying formalism, the prevailing attitude is to treat
them as individually inseparable entities.
According to this logic, a Feynman diagram either contributes to
$F(q^{2})$ in its entirety or it does not contribute at all.
This sort of logic is not part of the PT; certain diagrams, not relevant
to the definition of $F(q^{2})$ at first glance, contain pieces which cannot
be distinguished from the contributions of the regular graphs and must
therefore be included. It is precisely the inclusion of these contributions
which renders the answer g.i. and finite.\cite{Klako}
\eject
\subsection {Top magnetic dipole moment}

One of the most efficient ways to study
top quarks will be to pair-produce them in future very energetic $e^{+}e^{-}$
colliders, through the reaction $e^{+}e^{-}\rightarrow t\bar{t}$.
In general, the leptonic nature of the target allows for clean signals.
In addition, due to their large masses,
the produced top quarks are expected to
decay weakly
($t\bar{t} \rightarrow bW^{+}\bar{b} W^{-}$, with subsequent leptonic
decays of the $W$), before hadronization takes place; therefore
electroweak properties of the top can be studied in detail and QCD
corrections can be reliably evaluated in the context of perturbation theory,
when the energy of the collider is well above the threshold for $t\bar{t}$
production.

The standard method for extracting theoretical information out
of such an experiment is
to evaluate in the context of a specific gauge theory, such as the SM or its
extensions,
all Feynman graphs contributing to the
process $e^{+}e^{-}\rightarrow t\bar{t}$,
up to a given order in perturbation theory, compute the value of
an appropriately chosen observable, such as the
cross-section or the production rate,
and than compare it with the value obtained
experimentally.\cite{Hollik}
An alternative approach
is to to parametrize amplitudes in terms of form factors of
particles. The main motivation of such a method is to isolate
possible new physics
in a particular sub-amplitude, assuming that the rest
of the dynamics has already been successfully tested
in previous experiments.
Adopting the latter approach, Atwood and Soni~\cite{Atwood}
presented
a phenomenological analysis for determining the magnetic
and electric dipole moment form factors
of the top quark in upcoming  $e^{+}e^{-}\rightarrow t\bar{t}$
experiments.
Such form factors are defined through the
following Lorentz
decomposition of the $Vt\bar{t}$ vertex,
where V represents a boson (a $\gamma$ or $Z$ in our case)
coupled to the conserved leptonic current:
\begin{equation}
\Gamma_{\mu}^{V}
(q^2) = \gamma_{\mu}F_{1}^{V}(q^2)+ \Sm F_{2}^{V}(q^2)
+\gamma_{\mu}\gamma_{5} F_{3}^{V}(q^2)
+\Sm\gamma_{5} F_{4}^{V}(q^2)~,
\label{GeneralForm}
\end{equation}
with $q^2=s$ the Mandelstam variable associated to the squared energy of
the center of mass.
In the above decomposition, $F_{2}^{V}$ is the magnetic dipole moment
(MDM) and
$F_{4}^{V}$ is the electric dipole moment (EDM) form factor.
In particular, $F_{2}^{\gamma}$ defined at $q^2=0$ is the
usual definition of the anomalous magnetic moment.
In the case of the top quark
production, clearly $q^{2}\ge 4m_{t}^{2}$.
Within the SM the tree-level value for
both $F_{2}^{V}$ and $F_{4}^{V}$ is zero.
The upshot of the analysis~\cite {Atwood} was that the
dependence
of the differential cross section for
the reaction $e^{+}e^{-}\rightarrow t\bar{t}$
on the real and imaginary (absorptive) parts
of the MDM and EDM form factors, for an incoming photon or $Z$,
can be singled out
{\sl individually},
through a set of optimally chosen
physical observables.
The theoretical prediction for these observables is obtained by
calculating the tree-level
amplitude for $e^{+}e^{-}\rightarrow t\bar{t}$,
using
the $Vt\bar{t}$ vertices of Eq(\ref{GeneralForm}),
instead of the usual tree-level vertices.
Clearly, the effective vertex of Eq(\ref{GeneralForm}) can
only be used for tree-level computations, since its inclusion in loops
will give rise to non-renormalizable divergences.
The result of such a tree-level computation is
g.i., if one {\sl assumes} that
the quantities $F_{i}^{V}$ do {\sl not} depend explicitly on
the gauge-fixing parameter $\xi$.
Indeed, in that case
the only dependence on $\xi$ is proportional to the
longitudinal part of the $\gamma$ or $Z$ propagator, and therefore
vanishes, as long as the leptonic current is conserved
($m_{e}=0$).
The final answer is expressed in terms of
$F_{2}^{V}$ and $F_{4}^{V}$, which at this level are treated as free
parameters.
Comparison of these expressions with the
experimentally obtained quantities, can yield, after appropriate
fitting, the experimental values of $F_{2}^{V}$ and $F_{4}^{V}$.
Clearly, before any possible non-zero experimental
values for $F_{2}^{V}$ and $F_{4}^{V}$ can be attributed
to Physics beyond the SM, one
ought to first take into account the
contributions induced by quantum corrections from the SM.
So, $F_{2}^{V}$
becomes non-zero through one-loop
quantum corrections, whereas $F_{4}^{V}$,
which violates $CP$, receives its first non-vanishing contributions
at three loops
Such contributions are traditionally
extracted from the one-loop
corrections to the $\gamma t\bar{t}$ and $Zt\bar{t}$ vertices;
clearly the
resulting amplitude is of the desired form,~\cite{Atwood}
namely a bare $\gamma$ or $Z$ propagator multiplied by a vertex of the
form of Eq(\ref{GeneralForm}).
However,
as it was pointed out already in the classic paper by Fujikawa, Lee, and
Sanda,~\cite{Fujikawa}
off-shell form factors of fermions are in general gauge dependent
quantities.
In the context of the $R_{\xi}$ gauges, for example, a residual dependence
on the gauge-fixing parameter $\xi$ survives in the final expressions
of form factors,
when $q^{2}\not = 0$.
Obviously, in the case of $e^{+}e^{-}$ annihilation into heavy fermions,
the value of $q^{2}$ must be above the heavy fermion threshold
($q^{2}\ge 4m_{t}^{2}$, in our case). Consequently, the intermediate
photon and $Z$ are far off-shell, and therefore,
MDM and EDM form factors may in general be gauge-dependent and not
suitable for comparison with experiment.
This gauge dependence
was computed~\cite{Claudio} and turned out to be
numerically very strong; its presence distorts
not only the quantitative but also the qualitative
behavior of the answer. More specifically,
unphysical thresholds
 are introduced, and the
numerical dominance
of perturbative QCD, which is present in the gauge g.i.
treatment,
 is totally washed out. Of particular interest is the fact that the
popular unitary gauge
( the limit of the $R_{\xi}$ gauges as $\xi\rightarrow\infty$) gives
a completely wrong answer. This analysis indicates that the
gauge dependence established is a serious pathology and may
lead to erroneous conclusions.
 Applying the PT to
the case of the MDM form factor computed in a general
$R_{\xi}$ gauge means that
one has to identify vertex-like contributions contained in box
diagrams, which, when added to to the usual vertex graphs, render the result
$\xi$-independent.
Interestingly enough,
the g.i. answer so obtained
{\sl coincides} with the one derived
when {\sl only} the usual vertex graphs are consider
(without contributions from boxes),
but are evaluated in a special gauge, namely
the {\sl Feynman gauge} ($\xi=1$).
\eject
\subsection{The $S$, $T$, and $U$ parameters}

One of the most frequently used parametrizations of the leading
contributions of electroweak radiative corrections is in terms
of the $S$, $T$, and $U$ parameters.\cite{Peskin}
As was shown by Degrassi,
Kniehl, and Sirlin,\cite{STU}
in the context of the standard model,
these parameters
become infested with gauge-dependences, as soon as
the bosonic contributions
to the one-loop
self-energies are taken into account. In addition,
these quantities are in general ultraviolet divergent, unless
one happens to work within a very special class of gauges,
namely those satisfying the constraint
\begin{equation}
\xi_{W}= \xi_{\gamma}sin^{2}\theta
+ \xi_{Z}cos^{2}\theta~.
\end{equation}
The above shortcomings may be circumvented if one defines the
$S$, $T$, and $U$ parameters through the g.i. PT self-energies
$WW$, $ZZ$, $\gamma Z$, and $\gamma\gamma$.

\section{Anomalous gauge boson couplings}
A new and largely unexplored frontier on which the ongoing search for new
physics will soon focus is the study of the structure of the three-boson
couplings. In particular one expects to probe directly the
non-Abelian nature of the standard model at LEP2 (and NLC),
through the process $e^{+}e^{-}\rightarrow W^{+}W^{-}$.
A general parametrization of the trilinear gauge boson vertex for
on-shell $W$s and off-shell  $V=\gamma,~Z$ is
\bea
\Gamma_{\mu\alpha\beta}^{V}= & -ig_{V}~\Big[~
 ~f \left[~ 2g_{\alpha\beta}\Delta_{\mu}+ 4(g_{\alpha\mu}Q_{\beta}-
g_{\beta\mu}Q_{\alpha})~\right]
 +~ 2\Delta\kappa_{V}~(g_{\alpha\mu}Q_{\beta}-g_{\beta\mu}Q_{\alpha})
\nonumber \\
& ~~+~  4\frac{\Delta Q_V}{M_W^2}
(\Delta_{\mu}Q_{\alpha}Q_{\beta}-
\frac{1}{2}Q^{2}g_{\alpha\beta}\Delta_{\mu})
{}~~\Big]~ +~ ... ~~~,
\eea
with $g_{\eg}= gs$, $g_{Z}= gc$, where $g$ is the  $SU(2)$
gauge coupling, $s\equiv sin\theta_{W}$ and $c\equiv cos\theta_{W}$,
 and the ellipses denote omission of C, P, or T violating terms.
 The four-momenta $Q$ and $\Delta$
 are related to the incoming momenta $q$, $p_{1}$ and $p_{2}$ of
the gauge bosons $V,~W^-$and $W^+$ respectively, by
$q=2Q$, $p_{1}=\Delta -Q$ and $p_{2}=-\Delta - Q$.\cite{Bardeen}
The form factors $\Delta\kappa_{V}$ and $\Delta Q_{V}$,
also defined as
$
\Delta\kappa_{V}= \kappa_{V} + \lambda_{V} - 1
$ ,
and
$
 \Delta Q_{V}= -2\lambda_{V}
$,
are compatible with C, P,
and T invariance, and
are  related to the magnetic dipole moment $\mu_{W}$ and the electric
quadrupole moment $Q_{W}$, by the following expressions:
\be
\mu_{W} = \frac{e}{2M_{W}}(2+ \Delta\kappa_{\gamma})
{}~~\mbox{,}~~
Q_{W}= -\frac{e}{M^{2}_{W}}(1+\Delta\kappa_{\gamma}+\Delta Q_{\gamma})~.
\ee
In the context of the standard model,
 their canonical, tree level values are
 $f=1$ and  $\Delta\kappa_{V}=\Delta Q_{V}=0$.
To determine the radiative corrections to these quantities
one must cast the resulting one-loop expressions in the following form:
\be
\Gamma_{\mu\alpha\beta}^{V}= -ig_{V}[
 a_{1}^{V}g_{\alpha\beta}\Delta_{\mu}+ a_{2}^{V}(g_{\alpha\mu}Q_{\beta}-
g_{\beta\mu}Q_{\alpha})
 + a_{3}^{V}\Delta_{\mu}Q_{\alpha}Q_{\beta}]~,
\label{1loopParametrization}
\ee
where $a_{1}^{V}$, $a_{2}^{V}$, and $a_{3}^{V}$ are
 complicated functions of the
momentum transfer $Q^2$, and the masses of the particles
appearing in the loops.
It then follows that  $\Delta\kappa_{V}$ and $\Delta Q_{V}$
are given by the
following expressions:
\be
\Delta\kappa_{V}=\frac{1}{2}(a_{2}^{V}-2a_{1}^{V}-Q^{2}a_{3}^{V})
{}~\mbox{,}~~
\Delta Q_{V}= \frac{M^{2}_{W}}{4}a_{3}^{V}~.
\label{1loopdeltakappaandQ}
\ee
Calculating the one-loop expressions for $\Delta\kappa_{V}$ and
 $\Delta Q_{V}$ is a non-trivial task, both from
the technical and the conceptual point of view.
If one
calculates
just the Feynman diagrams contributing to the $\gamma W^{+}W^{-}$
vertex and then extracts from them the contributions to
$\Delta\kappa_{\gamma}$ and $\Delta Q_{\gamma}$,
one arrives at
expressions that are
plagued with several pathologies, gauge-dependence being one of them.
Indeed, even if the two W are considered to be on shell, since the incoming
photon is not, there is no {\sl a priori}
 reason why a g.i. answer
should emerge. In the context of the renormalizable $R_{\xi}$ gauges
the final answer depends on the
choice of the gauge fixing parameter $\xi$, which enters into the one-loop
calculations through the gauge-boson propagators
( W,Z,$\gamma$, and unphysical Higgs particles).
In addition, as shown by an explicit calculation
performed
in the Feynman gauge ($\xi=1$), the answer for
$\Delta\kappa_{\gamma}$
is {\sl infrared divergent} and
violates perturbative unitarity,
e.g. it grows
monotonically for $Q^2 \rightarrow \infty$.\cite{Lahanas}
All the above pathologies may be circumvented
if one adopts the PT.\cite{Kostas}
The application of the PT gives rise to
new expressions,
$\hat{\Delta}\kappa_{\gamma}$ and $\hat{\Delta} Q_{\gamma}$,
which are
gauge fixing parameter ($\xi$) independent,
ultraviolet {\sl and} infrared finite, and
well behaved for large momentum transfers $Q^{2}$.

Using carets to denote the g.i. one-loop contributions.
we have
\be
\hat{\Delta}\kappa_{\gamma} = \Delta\kappa_{\gamma}^{(\xi=1)}
+ \Delta\kappa_{\gamma}^{P}~,
\ee
and
\be
\hat{\Delta}Q_{\gamma} = \Delta Q_{\gamma}^{(\xi=1)}+  \Delta Q_{\gamma}^{P}
{}~.
\ee
where $\Delta Q_{\gamma}^{(\xi=1)}$ and $\Delta Q_{\gamma}^{(\xi=1)}$
are the contributions of the usual vertex diagrams in
the Feynman gauge,\cite{Lahanas} whereas
$\Delta Q_{\gamma}^P$ and $\Delta Q_{\gamma}^P$the analogous
contributions from the pinch parts.
A straightforward calculation yields:
\be
{\Delta\kappa}_{\gamma}^{P}= -\frac{1}{2} \frac{Q^{2}}{M^{2}_{W}}
\sum_{V} \frac{\alpha_{V}}{\pi} \int_{0}^{1}da \int_{0}^{1}(2tdt)
\frac{(at-1)}{L^{2}_{V}}~,
\ee
and
\be
{\Delta Q}_{\gamma}^{P} = 0 ~~~.
\ee
where
\be
L^{2}_{V} = t^{2}-t^{2}a(1-a)(\frac{4Q^{2}}{M^{2}_{W}}) +
 (1-t)\frac{M^{2}_{V}}{M^{2}_{W}}~.
\ee
We observe that $\Delta\kappa_{\gamma}^{P}$ contains an
infrared divergent term, stemming from the double
integral shown above, when $V=\gamma$. This term cancels exactly
against a similar infrared divergent piece
contained in $\Delta\kappa_{\gamma}^{(\xi = 1)}$,
thus rendering
$\hat{\Delta}\kappa_{\gamma}$ infrared finite.
After the infrared pieces have been cancelled, one notices that
the remaining contribution of $\Delta\kappa_{\gamma}^{P}$
decreases monotonically
as $Q^2 \rightarrow \pm \infty$; due to the difference
in relative signs this contribution cancels
asymptotically against the monotonically increasing
contribution from
$\Delta\kappa_{\gamma}^{(\xi = 1)}$.
Thus by including the pinch part the
unitarity of $\hat{\Delta}\kappa_{\gamma}$ is restored and
$\hat{\Delta}\kappa_{\gamma} \rightarrow 0$ for large values of $Q^2$.
It would be interesting to determine how these quantities could be
directly extracted from future $e^{+}e^{-}$ experiments.

\section {Recent developments}
In the previous sections we presented the theoretical motivation for
introducing the PT, and discussed
its spectacular success on issues related
to electroweak phenomenology.
Several important issues remain however open.
Most noticeably,
it is crucial to establish on much firmer ground the
physical significance of the PT amplitudes, address issues
of uniqueness, extend the PT beyond one-loop, and
explore the possibility of directly extracting the PT form
factors from future experiments.
Some of the above questions have been addressed
in a series of relatively recent papers.
When reviewing these most recent developments
we will give additional emphasis on conceptual rather
than technical issues.
In particular, we will discuss
the application of the PT in the context of the unitary gauge,
the connection between the PT and the background field method,
the large mass limit of the S-matrix, and the process
independence of the PT results.

\subsection {The unitary gauge}
Since the early days of spontaneously broken non-Abelian gauge theories,
the unitary gauge has been known to give rise to
renormalizable S-matrix elements, but to
Green's functions
that are non-renormalizable in the sense that their divergent parts cannot
be removed by the usual mass and field-renormalization
counterterms.\cite{SLee}
Even though this
situation may be considered acceptable from the
physical point of view,
the inability to define
renormalizable Green's functions has
always been a theoretical shortcoming of the unitary gauge.
For example, the self-energies, vertex and box diagrams are divergent,
and gauge-boson propagators cannot be consistently defined for arbitrary
values of $q^{2}$ beyond the tree-level.
The application of the PT to the unitary
gauge calculations~\cite{PaSi}
systematically reorganizes the one-loop S-matrix contributions into
kinematically distinct  pieces (propagators, vertices, boxes) that can
be renormalized with the usual counter-terms characteristic of a
renormalizable theory.
The aforementioned shortcomings
associated with the unitary gauge
are thus circumvented.
Furthermore,
the renormalizable amplitudes obtained in this fashion
are {\sl identical} to those calculated
in the $R_{\xi}$ gauges.\cite{D&S}

It should be emphasized that the above results,
are by no means obvious.
The point is that the unitary gauge can be obtained from the
$R_{\xi}$ gauges if the limit $\xi\rightarrow\infty$ is taken before
Feynman integrals are performed. Thus, there is no obvious guarantee
that when the PT is applied directly to the highly divergent amplitudes
characteristic of the unitary gauge calculations, it will lead to the same
$\xi$-independent self-energies, vertices, and boxes derived in the
$R_{\xi}$ framework.

\subsection{The connection with the background field method}

Recently, a connection between the
background field method (BFM)~\cite{Abbott}
and the S-matrix PT,\cite{BFMGermany} and
subsequently the intrinsic PT~\cite{BFMJapan}
has been advertised.
In particular, it was shown that when QCD is
quantized in the context of BFM,
the conventional $n$-point functions, calculated with the
BFM Feynman rules,
{\sl coincide} with those obtained via the PT, for the special
value $\xi_{Q}=1$ of the
gauge fixing parameter $\xi_{Q}$,
used to gauge fix the quantum field.
For any other value of
$\xi_{Q}$ the resulting expressions differ from those obtained
via the PT. However, the BFM  $n$-point functions, for any choice of
$\xi_{Q}$, satisfy exactly the same Ward identities
as the PT $n$-point functions (Eq(\ref{WI2}) for example).
Based on these observations, it was argued
\cite{BFMGermany} that the PT is but
a special
case of the BFM, and represent one out of an infinite number of
equivalent choices, parameterized by the values chosen for $\xi_{Q}$.
This misleading point of view originates from the erroneous
impression that in the context of the BFM
Green's functions should be rendered g.i. automatically.
So, the naive expectation was that Green's functions calculated
within the BFM should be
completely g.i., and identical to the
corresponding PT Green's functions.
Therefore, when at the end of the calculation it was realized that,
contrary to the initial expectation, a
residual dependence on $\xi_{Q}$ persists,
there was an attempt
to assign a physical significance
to this dependence.
In particular, it was argued~\cite{BFMGermany}
that the residual $\xi_{Q}$
dependence is a trade-off for the (presumably) intrinsic
arbitrariness of the PT in defining $n$-point functions.
There is no {\it a priori}
reason however, why the Green's functions of
the BFM should not be gauge-dependent;
indeed, the requirement of gauge-invariance with respect
to the background field does {\it not} imply
gauge-invariance with respect to the quantum field.
In particular, the choice $\xi_{Q}=1$ acquires a special meaning
in the context of the BFM, {\it only}
because it coincides with the
result of a g.i. calculation, namely that of the PT.

It should be emphasized
that
to the extend that the BFM $n$-point functions display a residual
(even though mild) $\xi_{Q}$-dependence, one still has to apply the PT
algorithm, in order to obtain a unique g.i. answer. So,
the PT results can be recovered
for {\it every} value of $\xi_{Q}$
as long
as one properly identifies the relevant pinch contributions concealed in
the rest of the graphs contributing to the $S$-matrix element.\cite{PaBFM}
These contributions vanish for $\xi_{Q}=1$, but are {\it non-vanishing}
for any other value of $\xi_{Q}$.
In the case of the gluon
self-energy, for example,
one has to identify the propagator-like parts of
boxes and vertex graphs and, according to the PT rules, append them
to the conventional self-energy expressions. After this
procedure is completed, a unique PT result for the self-energy emerges,
regardless of the gauge fixing procedure (BFM, $R_{\xi}$, light-cone, etc),
or the value of the gauge fixing parameter
($\xi_{Q}$, $\xi$, $n_{\mu}$, etc) used.
{}From the point of view of the PT,
there is no real conceptual difference
between a theory quantized in the $R_{\xi}$ gauge or in the BFM.
Indeed, in the PT framework the crucial quantity is the S-matrix,
whose uniqueness and gauge independence is
systematically exploited, in order to extract
g.i. sub-amplitudes.
Even though
these sub-amplitudes have not yet been
associated with specific physical
observables, there are several indications
supporting such a possibility.
As it was recently realized,~\cite{Martin} for example,
the PT expression for the gluon self-energy coincides with the
renormalized
static quark-antiquark potential, in the limit of very heavy
quark masses.
The BFM, regardless of any calculational simplifications
it may cause, is bound to give rise
to the same S-matrix elements, order by order in perturbation theory.
It is therefore not surprising that the application of the PT
gives exactly the same
answers in the BFM, as in any other gauge fixing procedure.
The difference between various gauge fixing procedures
is only operational.
{}From that point of view
one could alternatively say
that the Feynman gauge ($\xi_{Q}=1)$ in the BFM has the special
property (at least at one-loop) of giving zero total pinch
contribution. To see that we recall that
the main characteristics of the Feynman rules in the BFM
\cite{Abbott} are that the gauge fixing parameters for the
background (classical) and the quantum fields are different
($\xi_{C}$ and $\xi_{Q}$ respectively),
the three and four-gluon vertices are $\xi_{Q}$-dependent at tree-level,
and the couplings to the ghosts are modified (they are however
$\xi_{Q}$-independent).
In particular, the three-gluon vertex assumes the
form~\cite{BFMGermany}
(omitting a factor $if_{abc}$)
\begin{equation}
{\Gamma}_{\mu\nu\alpha}^{(0)}= (\frac{1-\xi_{Q}}{\xi_{Q}})
\Gamma^{P}_{\mu\nu\alpha} +
\Gamma^{F}_{\mu\nu\alpha}~,
\label{tHooft}
\end{equation}
where
$
\Gamma^{P}_{\mu\nu\alpha} = (q+k)_{\nu}g_{\mu\alpha}
 + k_{\mu}g_{\nu\alpha}$~ gives rise to pinch
parts, when contracted with $\gamma$ matrices,
whereas
$\Gamma^{F}_{\mu\nu\alpha} = 2q_{\mu}g_{\nu\alpha} -
 2q_{\nu}g_{\nu\alpha} - (2k+q)_{\alpha}g_{\mu\nu}$~
cannot pinch.
Clearly, it vanishes for $\xi_{Q}=1$, and so do the longitudinal
parts of the gluon propagators; therefore
pinching in this gauge is zero.

\subsection {The large mass limit of the S-matrix}
An important open question is
if the g.i. quantities extracted
via the PT correspond to
physical quantities.
Using Eq(\ref{RunnCoupl}), it is straightforward to
verify that,
the one-loop expression
for ${\hat{T}}_{1}$ is :
\begin{equation}
{\hat{T}}_{1}={\bar{u}}_{1}\gamma_{\mu}{u}_{1}
 \{\frac{g^{2}}
{q^{2}[1+bg^2\ln(\frac{-q^2}{\mu^2})]}\}
{\bar{u}}_{2}\gamma^{\mu}{u}_{2}~~,
\label{LaPr}
\end{equation}
where $u_{i}$ are the external quark spinors.
Thus, up to the kinematic factor $\frac{1}{q^{2}}$, the
r.h.s. of Eq(\ref{LaPr}) is the one-loop running coupling.
Equivalently,
 the expression of Eq(\ref{LaPr}) is the Fourier transform
of the
static quark-antiquark potential, in the limit of very heavy
quark masses.\cite{Bill}
Clearly, the quark-antiquark potential
 is a physical quantity, which, at least in principle,
can be extracted from experiment, or measured on the lattice.
In fact, as was recently realized,\cite{Martin}
when one computes the one-loop contribution
to the scattering amplitude $q\bar{q}\rightarrow q\bar{q}$
of quarks with mass $M$,
retaining leading terms in $\frac{q^{2}}{M^{2}}$,
 one arrives again at
the expression of Eq(\ref{LaPr}).
So in principle, one can extract the quantity of Eq(\ref{LaPr})
from a scattering process, in which the momentum transfer
$q^{2}$ is considerably larger than the QCD mass $\Lambda^{2}$,
so that perturbation theory will be reliable,
and, at the same time, significantly smaller than the
mass of the external quarks, so that
the sub-leading corrections
of order $O(\frac{q^{2}}{M^{2}})$
 can be safely neglected.
Top-quark
scattering, for example, could provide
a physical process, where the above requirements
are simultaneously met.
The above observations led to the conjecture
that the PT
expressions for the gluonic $n$-point functions
correspond
to the static potential of a system of $n$ heavy
quarks.\cite{PaBFM}

\subsection {Process independence of the pinch technique}
The most recent development addresses the issue of
process-independence of the PT results.
In particular, the g.i.
$n$-point functions
obtained by the application of the S-matrix PT do {\sl not}
depend on the particular process employed
(fermion + fermion $\rightarrow$
fermion + fermion,~ fermion + fermion $\rightarrow$ gluon + gluon,
{}~gluon + gluon $\rightarrow$ gluon + gluon, etc.),
and are in that
sense universal. This fact can be seen with an
explicit calculation, where
one can be convinced that the only
 quantities entering in the definition of the
g.i. self-energies are just the gauge group
structure constants; therefore, the only difference
from process to process
is the external group matrices associated with external-leg wave functions,
which are, of course, immaterial
to the definition of the things inside.
The fact that the PT gives rise to
{\it process-independent} results has been recently proved
by N.~J.~Watson
\cite{Watson} via detailed calculations for a wide variety
of cases.


\section{Acknowledgements}
The author is indebted to the Max Planck Institute of
M\"unich,
and especially to Bernd Kniehl, for the
warm hospitality extended to him during his visit.
The author thanks
N.~J.~Watson for many stimulating discussions.
This work was supported by the National Science Foundation
under Grant No.~PHY-9313781.
\eject

\end{document}